\documentclass[aps]{revtex4-2}
\usepackage[utf8]{inputenc}
\usepackage{bm}
\usepackage{afterpage}
\usepackage{graphicx}
\usepackage{xcolor}
\usepackage{amsmath}
\usepackage{mathpazo}
\usepackage[margin=15mm]{geometry}
\usepackage[font=small,labelfont=bf]{caption}
\usepackage{multirow}
\newcommand{\ee}{\mathrm{e}}

\begin{document}
\title{Feasibility of a Pulsed Ponderomotive Phase Plate for Electron Beams}
\author{K.A.H. van Leeuwen$^{1,*}$, W.J. Schaap$^1$, B. Buijsse$^2$, S. Borelli$^1$, S.T. Kempers$^1$, W. Verhoeven$^{1,2}$ and O.J. Luiten$^1$}
\address{$^1$Department of Applied Physics, Eindhoven University of Technology, PO Box 513, 5600MB Eindhoven, The Netherlands}
\address{$^2$Thermo Fisher Scientific, Achtseweg Noord 5, 5651 GG Eindhoven, The Netherlands}
\address{$^*$Author to whom any correspondence should be addressed.}
\date{\today}
\begin{abstract}We propose a scheme for constructing a phase plate for use in an ultrafast Zernike-type phase contrast electron microscope, based on the interaction of the electron beam with a strongly focused, high-power femtosecond laser pulse and a pulsed electron beam. Analytical expressions for the phase shift using the time-averaged ponderomotive potential and a paraxial approximation for the focused laser beam are presented, as well as more rigorous quasiclassical simulations based on the quantum phase integral along classical, relativistic electron trajectories in an accurate, non-paraxial description of the laser beam. The results are shown to agree well unless the laser beam is focused to a waist size below a wavelength. For realistic (off-the-shelf) laser parameters the optimum phase shift of $-\pi/2$ is shown to be achievable. When combined with RF-cavity based electron chopping and compression techniques to produce electron pulses, a femtosecond regime pulsed phase contrast microscope can be constructed. The feasibility and robustness of the scheme are further investigated using the simulations, leading to motivated choices for design parameters such as wavelength, focus size and polarization. 
\end{abstract}
\maketitle

\section{Introduction}
Just like optical microscopes, standard transmission electron microscopes have a problem with transparent samples that induce mainly phase and little amplitude variations in the transmitted beam. Especially biological samples fall into this category. With ideal imaging, the result is a zero to low contrast image on the detector in the image plane. For optical microscopes, the standard solution is to use a phase contrast microscope. In such a microscope an optical phase plate is used in the back focal plane of the microscope objective in such a way that it selectively shifts the phase of either the scattered or unscattered part of the light transmitted through the sample. The result is that the phase variations across the sample translate into intensity variations across the detector, thus strongly enhancing the contrast of the image. This technique has been developed by Zernike \cite{Zernike1934a,Zernike1934b,Zernike1938}, for which he was awarded the Nobel prize in 1953. The optimal shift of the phase plate in a Zernike-type phase contrast microscope is $\pm \pi/2$, the sign determining positive or negative contrast for a given phase imprint.

For a TEM, the same principle can be applied. However, making a well-controlled, durable and reliable phase plate for electrons is far more difficult for electrons than for light. A number of schemes have been proposed (see, e.g., \cite{Danev15635} and references therein). Phase plates that have been used in practical applications include thin carbon films with a small hole in the center through which the unscattered electrons pass \cite{DANEV2001243} and thin films where the strong, highly localized unscattered beam induces itself a local phase shift by structurally altering the thin film material (Volta phase plates) \cite{Danev15635}. However, there are problems with the reliability and stability of the holed thin films and as yet unexplained losses in the Volta phase plates \cite{BUIJSSE2020113079}.

Eliminating the thin films through which the electrons have to pass in most of the schemes would remedy many of the disadvantages. A strongly focused intense laser beam can be used as such a non-material phase plate, with the ponderomotive potential in the focus inducing the localized phase shift. In addition, a laser-based phase plate acts as a variable phase plate, with the magnitude of the phase shift determined by the laser intensity. This can be used to reconstruct the complete electron wavefront after passing through the sample, even for thick samples~\cite{VANDYCK2010571}.

The ponderomotive phase plate has been proposed by M\"uller {\it et al.} \cite{Mueller2010} and realized by the same group \cite{Schwartz2019,Turnbaugh2021}. The authors used a continuous laser and a near-concentric optical buildup cavity inside the vacuum in order to obtain the light intensity needed to produce sufficient phase shift for the electrons passing through the laser focus. They succeeded in achieving the optimal phase shift of $\pi/2$~rad for electrons passing through an antinode of the standing wave near the focus. 

As an alternative approach, we propose to use a high-power femtosecond pulsed laser in combination with a pulsed electron beam instead of a continuous wave laser with a buildup cavity. Although this restricts the phase contrast microscope to pulsed operation, limiting the scope of applicability, it can be achieved in a much simpler setup. Pulsed lasers of sufficient intensity are easily available commercially and modifications to existing microscopes are less intrusive. Furthermore, a pulsed phase plate can find many applications in the fast expanding area of ultrafast transmission electron microscopy. 

In order to evaluate the feasibility of this scheme and to provide data for making informed design choices, we first outline a simple model based on the time-averaged ponderomotive (quasi)potential and a focused Gaussian beam in the paraxial approximation. Using this model, the phase shift as well as the changes in the axial and transverse velocity components of the electrons can be evaluated analytically.

As a very tight laser focus and short laser pulses are needed to achieve the required phase shift, both the validity of using the paraxial approximation for the laser beam and of using only the time-averaged ponderomotive potential can be questioned. Therefore we next outline a much more rigorous approach in which the induced electron phase shifts are evaluated numerically on the basis of relativistic classical trajectory calculations using a quasi-classical path integral approach. For the trajectory calculations, the electromagnetic field of the strongly focused laser beam is modeled in full non-paraxial form using the angular spectrum representation method. This allows us not only to evaluate the phase shift and velocity changes in a more rigorous way, but also to look at the influence of the laser polarization (absent when using the ponderomotive potential).

\section{Analytical estimates\label{Sec:pondero_ana}}
In order to obtain analytical estimates we start with the time-averaged ponderomotive (quasi)potential for a charged particle in a linearly polarized electromagnetic field:
\begin{equation}
U_p=\frac{q^2 E_0^2}{4m_r\omega_0^2} \label{Eq:pondero_pot}
\end{equation}
with $q$ the charge of the particle, $m_r=m\gamma$ ($\gamma=1/\sqrt{1-|\bm{v}|^2/c^2}$ the Lorentz factor) its relativistic mass, $E_0$ the electric field amplitude and $\omega_0$ the frequency of the field. 

We assume a Gaussian laser beam in the paraxial approximation propagating along the z-axis, with a waist $w_0$ at $z=0$ and local width $w(z)$. The laser is pulsed with a Gaussian time envelope, duration (RMS power) $\tau$, pulse energy $\mathcal{W}$ and angular frequency $\omega_0$. The maximum of the laser pulse envelope arrives at time $t=0$ at $z=0$.  The electron is traveling in the positive x-direction with initial velocity $\bm{v}=(v_x,0,0)$ and is at the position $(0,y,z)$ at time $t=t_0$. Furthermore, we assume that the velocity of the electron remains approximately unchanged during the interaction. With these assumptions, the ponderomotive potential can be written
\begin{equation}
U_p(x,y,z,t)=\frac{e^2}{c\epsilon_0 m_r \omega_0^2}\,\,\frac{\mathcal{W}}{\pi^{\frac{3}{2}}w(z)^2\tau}\ee^{-\frac{(t-z/c)^2}{\tau^2}}\ee^{-2\frac{(v_x(t-t_0))^2+y^2}{w(z)^2}}. 	
\label{Eq:Pondero_pulse}
\end{equation}

\subsection{Phase shift\label{Subsec:phase_ana}}
Integrating the ponderomotive quasipotential over time results in an analytical expression for the phase shift:
\begin{align}
\Delta\varphi_p&=\frac{-1}{\hbar}\int_{-\infty}^{+\infty}{d\mkern-2mu t}U_p(v_x(t-t_0),y,z,t) \nonumber \\
&=\frac{-e^2}{\hbar c\epsilon_0 m_r \omega_0^2}\,\,\frac{\mathcal{W}}{\pi w(z)^2\tau}\frac{1}{\sqrt{\frac{1}{\tau^2}+\frac{2v_x^2}{w(z)^2}}}\ee^{-2\frac{y^2}{w(z)^2}}\ee^{-2\frac{\left(v_x(\frac{z}{c}-t_0)\right)^2}{2\tau^2v_x^2+w(z)^2}}
\label{Eq:Phaseshift}
\end{align}
Optimal alignment in time and space between laser pulse and electron ($t_0=0$, $y=0$, $z=0$) results in the maximum absolute value for the (negative) ponderomotive phase shift:
\begin{align}
\Delta\varphi_{\textrm{max}}&=\frac{e^2}{\hbar c
	\epsilon_0 m_r \omega_0^2}\,\,\frac{\mathcal{W}}{\pi w_0^2}\frac{\tau_t}{\tau_s}. \label{Eq:Maxshift_rel}
\end{align} 
In this expression we have defined the effective transit time of the electron through the laser waist as $\tau_t=w_0/\sqrt{2}v_x$. $\tau_s\equiv\sqrt{\tau^2+\tau_t^2}$ is the quadratic sum of laser pulse duration and transit time. 

For these expressions to be valid, the time averaging implicit in the definition of the ponderomotive quasipotential has to be justified. This requires both the pulse duration and the transit time to be much larger than the optical cycle time ($\tau \gg 1/\omega_0, \tau_t \gg 1/\omega_0$).

Eq.~\ref{Eq:Maxshift_rel} indicates that a small waist, large wavelength and large transit time increase the maximum achievable phase shift. However, these parameters are not independent: the transit time is proportional to the waist size, and the waist size can be expressed in terms of the wavelength and the numerical aperture or focusing parameter of the Gaussian beam, $\varepsilon=\lambda/\pi w_0$ . The latter parameter is, in a practical setup, limited by the numerical aperture of the focusing optics used. It is therefore useful to rewrite Eq.~\ref{Eq:Maxshift_rel} as:
\begin{align}
	\Delta\varphi_{\textrm{max}}&=\frac{e^2}{4\sqrt{2}\pi^2 \hbar c^3 \epsilon_0 m_r v_x} \frac{\lambda\mathcal{W}\, \varepsilon}{\tau_s}\label{Eq:Maxshift_rel2}
\end{align}
If we further assume that the laser pulse duration is much larger than the transit time ($\tau \gg \tau_t$), we can replace $\tau_s$ in this expression by $\tau$.

$v_x$ is generally fixed by the operating energy of the electron microscope: at 200~keV, $v_x \approx 2 \times 10^8\,\textrm{ms}^{-1}$. Thus, within the long pulse limit and at fixed laser pulse energy, the maximum phase shift is inversely proportional to the pulse duration, and proportional to the laser wavelength and the numerical aperture of the optics used. Thus, large phase shifts can more easily be obtained using a large wavelength laser. However, the effective size of the phase plate also increases linearly with the wavelength, and a large size results in a phase contrast image that is limited to high spatial frequency.  
 
In the derivation the assumption is made that relativistic effects are fully accounted for by the replacement of the rest mass of the electron by the relativistic mass in Eq.~\ref{Eq:pondero_pot}, which may constitute a simplification as well.

\subsection{Velocity changes\label{Subsec:delv_ana}}
Imperfect alignment in time and space (unavoidable with a finite sized electron beam with pulses of finite duration) will not only decrease the phase shift according to Eq.~\ref{Eq:Phaseshift}, but can also effect the axial and transverse velocity components of the electron. These can effect the image formation in the electron microscope and therefore have to be considered when evaluating the feasibility of the scheme.

In the same approximation as used in the previous section to derive an estimate of the phase shift (i.e., using a paraxial Gaussian beam and the time-averaged ponderomotive potential of Eq.~\ref{Eq:Pondero_pulse}), these velocity changes can be analytically evaluated. Assuming that the electron remains very close to the $z=0$ focus plane, we replace $w(z)$ with $w_0$.  The force on the electron from the gradient of the ponderomotive potential can then be written:
\begin{align}
\bm{F}=-\bm{\nabla}U_p=&\frac{-e^2}{c\epsilon_0 m_r \omega_0^2}\,\,\frac{\mathcal{W}}{\pi w(z)^2\tau} \ee^{-\frac{(t-t_0+z/c)^2}{\tau^2}} \ee^{-2\frac{x^2+y^2}{w_0^2}} 
\left[\frac{4x}{w_0^2} \,\hat{x} + \frac{4y}{w_0^2} \,\hat{y} - \frac{2(t-t_0+z/c)}{c\tau^2} \, \hat{z}\right]. \label{Eq:PonderomotiveForce}
\end{align}

Writing the initial axial velocity $v_{x,\textrm{initial}} \equiv u$, and assuming the change in axial velocity to be small ($\Delta v_x\ll u$), we can put $x=u(t-t_0)$. Furthermore, we neglect the lateral shifts in the electron trajectory and replace $y$ and $z$ by their initial values $y_0$ and $z_0$. Integrating the force over time then gives the momentum change:
\begin{align}
\Delta\bm{p}&=\int_{-\infty}^{+\infty}\bm{F}\,dt = \frac{2 e^2}{c\epsilon_0 m_r \omega_0^2}\,\,\frac{\mathcal{W}}{\pi w_0^2}\frac{\tau_t}{\tau_s} \ee^{-\frac{(t_0-z_0/c)^2}{\tau_s^2}}  \ee^{-2\frac{y_0^2}{w_0^2}}
\left[\frac{1}{u\tau_s^2} \, (t_0-z_0/c)\, \hat{x} + \frac{1}{u^2\tau_t^2}\,y_0\, \hat{y} + \frac{1}{c\tau_s^2} \, (t_0-z_0/c)\,\hat {z}\right].
\label{Eq:Delta_p}
\end{align}
The momentum changes in the $x$-- and $z$--direction show a dispersive dependence on the arrival time of the maximum of the laser pulse at $z=z_0$. This results in a ``chirp'' of the axial velocity and an angular sweep of the beam in the $z$--axis over the duration of the electron pulse. In the $y$--direction the laser focus acts like a focusing lens.

\section{Rigorous approach \label{Sec:pondero_semi}}
In this section we outline a more general and rigorous procedure to calculate the ponderomotive phase shift on the basis of relativistic, classical electron trajectories calculated with and without the presence of a strongly focused laser beam without assuming the paraxial approximation. 
\subsection{Quasiclassical phase shift \label{Subsec:phase_quasi}}
The velocity changes follow directly from the trajectory calculations. In order to evaluate the phase shift, we use the quasiclassical approximation. In this approximation, valid as long as the electronic de Broglie wavelength is much smaller than the optical wavelength, the quantum mechanical phase $\phi$ is given by the action integral (integral of the Lagrangian $\mathcal{L}$ over time) along the classical path $P$ divided by $\hbar$:
\begin{equation}
\phi=\frac{1}{\hbar}\int_P \mathcal{L} \, dt \label{Eq:Phase_int_L}
\end{equation}
The relativistic Langrangian of a charged particle in an electromagnetic field is given by
\begin{equation}
\mathcal{L}=-\frac{mc^2}{\gamma}+q\bm{v}\cdot\bm{A}-q\Phi \label{Eq:Lagrange_rel}
\end{equation}
with $q$ the charge, $\gamma=1/\sqrt{1-|\bm{v}|^2/c^2}$ the Lorentz factor, $\bm{A}$ the vector potential and $\Phi$ the scalar potential. 

Using the classical--quantum correspondences
\begin{equation}
\hbar\bm{k}\Leftrightarrow\bm{p}
\end{equation}
\begin{equation}
\hbar \omega \Leftrightarrow H = \frac{(\bm{p}-q\bm{A})^2}{2m}+q\Phi 
\end{equation}
with $H$ the Hamiltonian and $\bm{p}=\gamma m \bm{v}+q\bm{A}$ the canonical momentum of the particle,  Eq.~\ref{Eq:Phase_int_L} can be rewritten as 
\begin{equation}
\phi = \int_P \left(\bm{k}\cdot d\bm{r}-\omega\, dt\right)=\int_P dt \left(\bm{k}\cdot \bm{v}-\omega\right). \label{Eq:phase_int}
\end{equation}
This form is easily interpreted in the quasiclassical approximation as the sum of the spatial and temporal contributions to the accumulated phase of the quantum mechanical wave along the classical path.

Eq.\ \ref{Eq:Phase_int_L} can be used to evaluate the phase shift. However, we have to be careful: the phase shift that we wish to calculate is defined as the difference between the phase with and without the field present. The two trajectories have to merge asymptotically in order to be able to assign a unique, meaningful value to the induced phase shift. 


In the specific case we are considering here, i.e., an electron beam intersecting a Gaussian or near-Gaussian laser beam, considering paths that begin and end outside the range of the laser beam (thus, $\bm{A}$ and $\Phi$ are zero at the beginning and end of the classical paths), and in the relevant parameter range considered in this paper, the final values for the momentum and hence for $\bm{k}$ ($\bm{k}_f$) in both trajectories (with and without laser beam) are equal to a very good approximation. The impact of small deviations will be discussed in Section~\ref{Sec:velocity_changes}.
Depending on the polarization of the light, there may be a translation of the final electron trajectory. This translation is much smaller than the transverse coherence length and hence negligible.

For both situations we can then evaluate the phase integral over a given time interval and determine the difference. However, we have to realize that even when the conditions mentioned above are fulfilled, and using the same initial position and velocity for the particle in both situations, the final positions at the end of the time interval will generally not be the same. The phase shift that we seek has to be defined at the same position and time. This implies that an extra contribution to the phase shift is needed to correct for this.  

Expressed in terms of canonical momentum instead of velocity, the phase integral evaluates to:
\begin{equation}
\phi = \frac{1}{\hbar} \int_P dt \left( \frac{-mc^2}{\gamma}+\frac{q\bm{p}\cdot \bm{A}}{\gamma m} - \frac{q^2\bm{A}\cdot \bm{A}}{\gamma m} -q\Phi \right) . \label{Eq:phase_rel_p}
\end{equation}

With $P_A$ and $P_N$ representing the classical paths of the particle with the electromagnetic field present and absent respectively, the phase shift can be written:
\begin{equation}
\Delta\phi = \frac{1}{\hbar}\left[\int_{P_A} dt \left(\frac{-mc^2}{\gamma}+\frac{q\bm{p}\cdot \bm{A}}{\gamma m} - \frac{q^2\bm{A}\cdot \bm{A}}{\gamma m} - q\Phi\right) - \int_{P_N} dt \left(\frac{-m c^2}{\gamma}\right)\right]-\frac{\bm{p}_f}{\hbar}\cdot \Delta\bm{r} .
\label{Eq:Phasedif_rel}
\end{equation}

The last term in Eq.~\ref{Eq:Phasedif_rel} represents the contribution to the final phase shift associated with the possible delay (or advance) of the particle as discussed above. In this term, $\bm{p}_f$ is the final value of the particle momentum (assumed to be equal for both trajectories) and $\Delta\bm{r}=\bm{r}_f^{P_A}-\bm{r}_f^{P_N}$ is the difference in final positions.

Inserting $q=-e$ for electrons, and regrouping the terms, we finally arrive at:
\begin{equation}
\Delta\phi = \frac{-mc^2}{\hbar}\left[\int_{P_A} dt \frac{1}{\gamma}-\int_{P_N} dt \frac{1}{\gamma}\right]-\frac{e}{\hbar}\left[\int_{P_A} dt \left(\frac{\bm{p}\cdot \bm{A}}{\gamma m} + \frac{e\bm{A}\cdot \bm{A}}{\gamma m}-\Phi\right)\right]-\frac{\bm{p}_f}{\hbar}\cdot \Delta\bm{r} .
\label{Eq:Phasedif_rel_f}
\end{equation}

In the non-relativistic limit, Eq.~\ref{Eq:Phasedif_rel_f} simplifies to:
\begin{equation}
\Delta\phi^\mathit{nr} = \frac{1}{2m\hbar}\left[\int_{P_A} dt \left(\bm{p}\cdot\bm{p}-e^2\bm{A}\cdot\bm{A} - me\Phi \right) - \int_{P_N} dt \left(\bm{p}\cdot\bm{p}\right)\right]-\frac{\bm{p}_f}{\hbar}\cdot \Delta\bm{r}
\label{Eq:Phasedif_nonrel_f}
\end{equation}
In the non-relativistic expression we recognize in the $\bm{A}\cdot\bm{A}$ term the time integral of the effective ponderomotive potential $e^2 E_0^2/4m\omega_\textrm{laser}^2$ and expect this to be the dominant term. In the relativistic expression, this is less obvious.

\subsection{Laser beam description beyond the paraxial approximation \label{Subsec:beyond_parax}}
\begin{figure}[h]
	\centering
	\includegraphics[width=100mm]{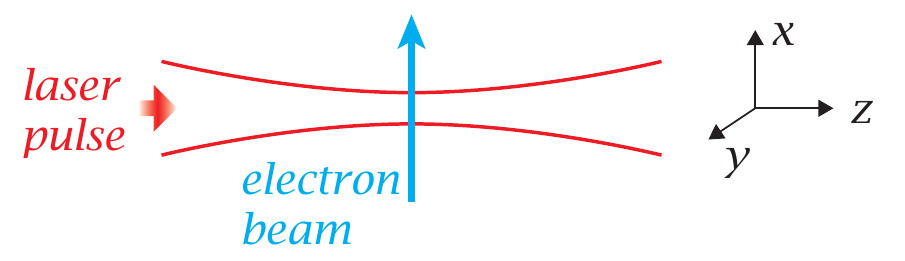}
	\caption{Geometry considered: the electron beam propagates along the $x$-axis and the laser beam along the $z$-axis. The laser beam is linearly polarized either along the $x$-axis (axial polarization) or along the $y$-axis (perpendicular polarization).}
	\label{Fig:Geometry}
\end{figure}

The basic geometry of the phase plate, with an electron crossing a linearly polarized focused laser beam, is shown in Figure~\ref{Fig:Geometry}. The laser beam propagates along the $z$--axis and is polarized either along the $x$-- or the $y$--axis, while the electrons travel along the $x$-axis.

Optimally, the size of the phase plate in a Zernike type electron microscope matches the size of the central spot in the back focal plane that is produced by electrons that are not diffracted by the sample. As this size (typically from a few tens to a few hundreds nanometer) remains below optical wavelengths in the visible regime, the laser beam should be focused to as small a spot as possible. Furthermore, Eq.~\ref{Eq:Maxshift_rel} indicates that, at fixed pulse energy, the achievable phase shift increases inversely proportional to the waist size when the transit time $\tau_t$ is smaller than the laser pulse duration (e.g., for an electron energy of 200~keV and a 30~fs laser pulse duration, $w_0$ should be below $4\,\mu$m). This also favors tight focusing. 

With a waist size at the focus comparable to the wavelength of the light, the paraxial approximation of a Gaussian laser beam is no longer valid. Fields and potentials in the waist plane cannot all have a Gaussian transverse profile, and field components along the propagation axis of the laser beam are non-zero. This affects the electron trajectories in the interaction region and hence may change the phase shift (Eq.~\ref{Eq:Phasedif_rel_f}). Thus, a more accurate description of the focused laser beam is required.

For this, we model the laser beam using the well-known method of angular spectrum representation of plane waves (Ref.~\cite{Ratcliffe1956,Sherman1982}). This method provides an accurate description of a laser beam with an arbitrary focus size. To apply the method, a choice has to be made for the profile of fields and potentials in the waist plane ($z=0$). Different choices have been treated in the works presented by, e.g., Quesnel and Mora \cite{Quesnel1998} and Sepke and Umstadter \cite{Sepke2006}. In our approach, we cover four possible choices:
\begin{itemize}
	\item A Gaussian distribution of the $x$-component of the electric field $E_x$ with zero $E_y$. This leads to a not-quite-Gaussian distribution of $B_y$ with non-zero $B_x$, as well as to non-zero $E_z$ and $B_z$.
	\item A Gaussian distribution of $B_y$ with zero $B_x$. This leads to a not-quite-Gaussian distribution of the $E_x$ with non-zero $E_y$-component, as well as to non-zero $E_z$ and $B_z$.
	\item The symmetrized choice, consisting of the average of the previous two choices.
	\item A linearly polarized Gaussian vector potential along the $x$-axis with no axial component. This leads to a not-quite-Gaussian $E_x$ and $B_y$, zero $B_x$, and non-zero $E_z$ and $B_z$. 
\end{itemize}
All other components of the fields and potentials at $z=0$, as well as all components for $z\neq0$, then follow from Maxwell's equations. As in Ref.~\cite{Quesnel1998,Sepke2006}, the circular symmetry of the chosen Gaussian field or potential component in the waist plane allows all final fields and potentials to be expressed in terms of a restricted number of one-dimensional integrals over (essentially) the diffraction angle. 

We directly evaluate these integrals numerically instead of using a series expansion in powers of the focusing parameter \cite{Quesnel1998} or a Fourier-Gegenbauer expansion \cite{Sepke2006}. This approach is simple and flexible. Moreover, it avoids convergence problems at either very small or large focus sizes, and includes the evanescent wave components of the fields. It is sufficiently fast for our purposes.  

The time dependence of the laser pulse is modeled as a simple Gaussian envelope traveling along the $z$-axis. Thus the effects of the wavefront curvature on the off-axis time dependence of the pulse and the wavelength dependence of the diffraction within the bandwidth of the pulse are not taken into account. This is justified because we consider electrons crossing the laser beam very close to the focus (i.e., within a small fraction of the Rayleigh length), and laser pulses with a duration, albeit short, still much longer than a single optical period. 

The laser pulse is thus represented according to one of the field configuration options by specifying a Gaussian field or potential component in the $z=0$ plane, e.g., for the first option (Gaussian $E_x$, in complex notation):
\begin{align}
E_x(x,y,0,t)&=E_0\,\, e^{-\frac{x^2+y^2}{w_0^2}} e^{\frac{(t-\frac{z}{c})^2}{2\tau^2}} e^{i \left(\omega_0 t+\phi_0\right)}\nonumber\\
E_y(x,y,0,t)&=0,
\end{align}
after which the other components are evaluated numerically.

In the paraxial limit, the full expressions for the fields and potentials simplify to:
\begin{align}
\bm{E}(x,y,z,t)&=E_0 \frac{w_0}{w(z)} e^{-\frac{x^2+y^2}{w(z)^2}} e^{\frac{(t-\frac{z}{c})^2}{2\tau^2}} e^{i \left(\omega_0 t-k_0 z-\phi_\textrm{G}(z)-\frac{k(x^2+y^2)}{2R(z)}+\phi_0\right)}\hat{x}\nonumber\\
\bm{B}(x,y,z,t)&=\frac{E_0}{c} \frac{w_0}{w(z)} e^{-\frac{x^2+y^2}{w(z)^2}} e^{\frac{(t-\frac{z}{c})^2}{2\tau^2}} e^{i \left(\omega_0 t-k_0 z-\phi_\textrm{G}(z)-\frac{k(x^2+y^2)}{2R(z)}+\phi_0\right)}\hat{y}\nonumber\\
\bm{A}(x,y,z,t)&=-i\frac{E_0}{\omega_0} \frac{w_0}{w(z)} e^{-\frac{x^2+y^2}{w(z)^2}} e^{\frac{(t-\frac{z}{c})^2}{2\tau^2}} e^{i \left(\omega_0 t-k_0 z-\phi_\textrm{G}(z)-\frac{k(x^2+y^2)}{2R(z)}+\phi_0\right)}\hat{x}\nonumber\\
\Phi(x,y,z,t)&=0 
\end{align}
with $w_0$ the minimum waist size, $\omega_0$ the angular frequency of the light, $k_0$ the wavenumber, $L = k_0 w_0^2/2$ the Rayleigh length,  $w(z) = w_0 \sqrt{1+(z/L)^2}$ the local waist size, $R(z) = (L^2+z^2)/z$ the wavefront curvature, and $\phi_\textrm{G}(z) = -\arctan{(z/L)}$ the Gouy phase. The phase of the field at $t=0$, $z=0$ is $\phi_0$.

An option is present in the simulation code allowing the field to be calculated according to each of the four "non-paraxial" choices for the field configuration, as well for the paraxial approximation. 

Evanescent wave contributions that become relevant at sub-wavelength waist size are included in this approach. For $w_0\ge\lambda$, they can be neglected. 
We assume that no static fields are present in the interaction region. However, the potentials are chosen to satisfy the Lorenz gauge, resulting in a time-dependent scalar potential related to the vector potential by $\Phi(t)=-c^2\bm{\nabla}\cdot\bm{A}$. 

Theory and detailed results of the field calculations will be presented in a forthcoming article.

\subsection{Trajectory calculations}
Using the approach outlined above to evaluate the fields for the chosen configuration option, polarization, waist size and pulse duration, the electron trajectories are calculated by numerical integration of the relativistic equations of motion:
\begin{align}
\frac{d\bm{p}_\textit{kin}}{dt}&=-e(\bm{E}+\frac{\bm{p}_\textit{kin}}{\gamma m}\times\bm{B}) \\
\frac{d\bm{r}}{dt}&=\frac{\bm{p}_\textit{kin}}{\gamma m}
\end{align}
where the kinetic momentum is related to the canonical momentum by $\bm{p}_\textit{kin}=\bm{p}+e\bm{A}$.

The field calculation proper always assumes polarization along the $x$--axis. Polarization along the $y$-- or $z$--axis is included by a coordinate transformation of the initial electron trajectory, such that the field becomes $x$--polarized, followed by a back transformation after the trajectory calculation.

The maximum intensity of the laser pulse arrives at the $z=0$ plane at $t=0$. As discussed before, off-axis corrections to the propagation of the pulse envelope due to the wavefront curvature and the wavelength dependence of the diffraction within the bandwidth of the pulse have not been considered. Neglecting the off-axis pulse envelope corrections is justified because only electrons entering the light field close to the focus are considered, and the motion in the $z$--direction of the electron induced by the interaction is, even in very strong light fields, always much smaller than an optical wavelength. Neglecting the wavelength dependence of the diffraction is justified because pulse durations $> 50 \,\textrm{fs}$ and wavelengths in the optical/near infrared regime are considered, such that $\tau >75 /\omega_0$.  

Using the calculated trajectories and the vector and scalar potential along these trajectories, the phase shift contributions of Eq.~\ref{Eq:Phasedif_rel_f} are evaluated. The velocity changes of the electron follow directly from the trajectory calculations.

\section{Feasibility, analytical \label{Sec:Feasibility_ana}}
\subsection{Required pulse energy \label{Sec:energy_needed}}
We will use the analytical expressions for the maximum phase shift (Eqs.~\ref{Eq:Maxshift_rel}--\ref{Eq:Maxshift_rel2}) to get an idea about the required parameters to achieve the targeted $-\frac{\pi}{2}$ phase shift. From these equations, we see that choosing a small beam waist, short laser pulse, low electron beam energy and long wavelength decrease the required laser pulse energy. For now, we will reduce the parameter space by assuming a typical $200\,\textrm{keV}$ electron beam energy and either $100\,\textrm{fs}$ or $300\,\textrm{fs}$ laser pulse duration. Although shorter laser pulses are feasible, we have to realize that, in order to achieve approximately the same phase shift for all (unscattered) electrons in the pulsed electron beam, 
the electron pulse has to be significantly shorter than the laser pulse (as well as perfectly synchronized with it). Using RF-cavity based electron beam chopping and compression techniques~\cite{Oudheusden2010,Pasmans2013,Franssen2017,Toonen2019,Rajabi2021}, $30\,\textrm{fs}$ pulses seem feasible, which would enable the use of a laser pulse with a width of $100\,\textrm{fs}$ or more. The $300\,\textrm{fs}$ pulse allows for longer electron pulses and puts less extreme demands on the synchronization.

\begin{table}[b]
	\begin{tabular}{|c|c|c|c|c|c|c|}
		\hline
		\multicolumn{3}{|c|}{} & \multicolumn{2}{|c|}{$\tau=100\,\textrm{fs}$}& \multicolumn{2}{|c|}{$\tau=300\,\textrm{fs}$}\\
		\hline
		$\lambda$ (nm) & $w_0/\lambda$ & $\varepsilon$ & $\mathcal{W}_n$ (nJ) &
		$R_{\textrm{1W}}$ (MHz)& $\mathcal{W}_n$ (nJ) &
		$R_{\textrm{1W}}$ (MHz)\\
		\hline 
		800 & 0.5 & 0.637 & 4.46 & 224 & 13.4 & 75 \\
		800 & 1   & 0.318 & 8.92 & 112 & 26.8 & 37 \\
		800 & 2   & 0.159 & 17.86 & 56 & 53.5 & 19 \\
		1500 & 0.5 & 0.637 & 2.38 & 420 & 7.13 & 140 \\
		1500 & 1   & 0.318 & 4.76 & 210 & 14.3 & 70 \\
		1500 & 2   & 0.159 & 9.56 & 105 & 28.6 & 35 \\
		\hline
	\end{tabular}
	\caption{Pulse energy $\mathcal{W}_n$ needed to achieve $\Delta \varphi_{\textrm{max}}=-\pi/2$ phase shift laser pulse durations $\tau=100\,\textrm{fs}$ and $\tau=300\,\textrm{fs}$ for different wavelengths and focusing parameters.  Pulse repetition rates leading to a fixed average power of 1~W, $R_{\textrm{1W}}$, are shown as well.}
	\label{Tab:Energy_needed}
\end{table}

At $200\,\textrm{keV}$ energy, the transit time exceeds the $100\,\textrm{fs}$ pulse duration for $w_0 > 29.5\,\mu\mathrm{m}$. Thus, for a tight focus and visible or near infrared wavelengths, we are well in the long pulse limit and can replace $\tau_s$ by $\tau$. The phase shift according to Eq.~\ref{Eq:Maxshift_rel2} is therefore proportional to the pulse energy, the wavelength and the focusing parameter, and inversely proportional to the pulse duration.

Table~\ref{Tab:Energy_needed} shows the required pulse energy to achieve $-\frac{\pi}{2}$ phase shift for two wavelengths and three focusing parameters. 
The pulse energies shown in the table are achievable with a variety of fs pulsed laser systems. However, with the average power of the system limited to a specific value, this determines a maximum pulse repetition rate. For an average power limited to 1~W, the associated repetition rates $R_{\textrm{1W}}$ are shown in the table as well.

\subsection{Velocity changes \label{Sec:velocity_changes}}
From the analytical expression for the momentum change (Eq.~\ref{Eq:Delta_p}) we can derive the changes in the final velocity $x$--, $y$-- and $z$--components induced by the interaction. We will assume $z=0$, and rewrite the expression to show the connection between the velocity changes and the ponderomotive phase shift:
\begin{align}
	\Delta v_x &=\frac{\Delta p_x}{\gamma^2 m_r} =\frac{2\hbar}{m}\Delta \varphi_p \frac{1}{\gamma^3 u \tau_s^2} t_0 \nonumber\\
	\Delta v_y &=\frac{\Delta p_y}{m_r} \,\,\,=\frac{2\hbar}{m}\Delta \varphi_p \frac{1}{\gamma u^2 \tau_t^2} y_0 	\label{Eq:Del_v_ana}\\
	\Delta v_z &=\frac{\Delta p_x}{m_r} \,\,\,=\frac{2\hbar}{m}\Delta \varphi_p \frac{1}{\gamma c \tau_s^2} t_0 \nonumber
\end{align}

With the parameters from Table~\ref{Tab:Energy_needed}, we are well within the long pulse limit and we can replace $\tau_s$ by $\tau$ in these expressions. As a consequence, the expression for the change in the axial velocity $\Delta v_x$ and in the velocity along the laser beam direction $\Delta v_z$ become independent of the laser wavelength and the waist size, provided that the pulse energy is chosen to produce a maximum phase shift of $-\frac{\pi}{2}$. With $\tau_t$ proportional to the waist size, deflections in the $y$--direction increase with decreasing waist size, but are independent of $\tau$. Table~\ref{Tab:Del_v_ana} shows, for the same parameters as in Table~\ref{Tab:Energy_needed}, the slopes $\partial v_x/\partial t_0$, $\partial v_y/\partial y_0$ and $\partial v_z/\partial t_0$ (at $t_0 = 0, y_0 = 0$) as well as absolute values of the extrema $|\Delta v_x^\textrm{max}|$, $|\Delta v_z^\textrm{max}|$ (at $|t_0|=\tau/\sqrt{2}, y_0=0$) and $|\Delta v_y^\textrm{max}|$ (at $t_0=0,|y_0|=w_0/2$).

\begin{table}[h]
\begin{tabular}{|c|c|c|c|c|c|c|c|c|c|c|c|c|c|}
	\hline
	\multicolumn{2}{|c|}{} & \multicolumn{6}{|c|}{$\tau=100\,\textrm{fs}$}& \multicolumn{6}{|c|}{$\tau=300\,\textrm{fs}$}\\
	\hline
	$\lambda$ & $w_0/\lambda$ & 
	$\frac{\partial v_x}{\partial t_0}$ & $|\Delta v_x^\textrm{max}|$ &
	$\frac{\partial v_y}{\partial y_0}$ & $|\Delta v_y^\textrm{max}|$ &
	$\frac{\partial v_z}{\partial t_0}$ & $|\Delta v_z^\textrm{max}|$ &
	$\frac{\partial v_x}{\partial t_0}$ & $|\Delta v_x^\textrm{max}|$ &
	$\frac{\partial v_y}{\partial y_0}$ & $|\Delta v_y^\textrm{max}|$ &
	$\frac{\partial v_z}{\partial t_0}$ & $|\Delta v_z^\textrm{max}|$\\
	\scriptsize{(nm)} & & \scriptsize{($10^{12}$m/s$^2$)} & \scriptsize{(m/s)} & \scriptsize{($10^9$/s)} & \scriptsize{(m/s)}& \scriptsize{($10^{12}$m/s$^2$)} & \scriptsize{(m/s)} & \scriptsize{($10^{12}$m/s$^2$)} & \scriptsize{(m/s)} & \scriptsize{($10^9$/s)} & \scriptsize{(m/s)} & \scriptsize{($10^{12}$m/s$^2$)} & \scriptsize{(m/s)}\\
	\hline 
	\multirow{3}*{800}  & 0.5 & \multirow{6}*{64.8} & \multirow{6}*{2.78} & 3.27 & 396 & \multirow{6}*{87.2} & \multirow{6}*{3.74} & \multirow{6}*{7.20}   & \multirow{6}*{0.93}     & 3.27 & 396 & \multirow{6}*{9.7} & \multirow{6}*{1.25} \\
	                    & 1   &                     &                     & 0.82 & 198 & & & & & 0.82 & 198 & & \\
	                    & 2   &                     &                     & 0.20 &  99 & & & & & 0.20 &  99 & & \\
	\cline{1-2} \cline{5-6} \cline{11-12}
	\multirow{3}*{1500} & 0.5 &                     &                     & 0.93 & 211 & & & & & 0.93 & 211 & &      \\
	                    & 1   &                     &                     & 0.23 & 106 & & & & & 0.23 & 106 & & \\
	                    & 2   &                     &                     & 0.06 & 53  & & & & & 0.06 & 53  & & \\
	\hline
\end{tabular}
\caption{Velocity changes induced by laser pulses for different wavelengths, focusing parameters and pulse duration. The electron beam energy is 200~keV. For each entry, the pulse energy is set to the value needed to achieve $\Delta \varphi_{\textrm{max}}=-\pi/2$ phase shift (see Table~\ref{Tab:Energy_needed}). $\tau_s$ is replaced by $\tau$ in Eqs.~\ref{Eq:Del_v_ana}.}
\label{Tab:Del_v_ana}
\end{table}

Table~\ref{Tab:Energy_needed} shows that a tight focus and a long wavelength decrease the power requirements for the laser. Furthermore, the focus size determines the cutoff value for the correctly imaged spatial frequencies (the smaller the focus, the better the contrast for low spatial frequency). According to the data from Table~\ref{Tab:Del_v_ana}, the only reason not to go for the tightest possible focus (except for the experimental difficulties) could be the (quadratic) increase in $\partial v_y/\partial y_0$ with decreasing waist size. 

A large value of $\partial v_y/\partial y_0$ corresponds to a strong focusing action for the electrons passing through the interaction region with the laser in the $y$--direction. However, this `lens action' for the electrons near the laser focus is intrinsic to any spatially inhomogeneous phase plate (as required in a Zernike microscope). 

In terms of Fourier optics, it is fully taken into account by the two-dimensional electron wave transfer function, of which the phase factor is equal to the phase shift distribution derived for the ponderomotive phase plate (together with contributions from aberrations).
The time dependent velocity changes in the $x$-- and $z$--direction have to be considered carefully. Even with perfect synchronization between the electron pulse and the laser pulse, the finite duration of the electron pulse translates into an angular sweep of the electrons passing through the focus and a similar chirp in the axial velocity. This effect does not occur, e.g., for the interaction with a focused CW laser. 

The angular sweep in the $z$--direction and the finite electron pulse duration can be translated in a spatial sweep of the position on the object that is imaged. It is not a large effect: for a 100~fs laser pulse, the maximum deflection angle at 200~keV electron energy is 18~nrad. With a typical effective focal length of 3~mm, this translates into an shift of 0.05~nm on the object plane. When placing the wave plate in a secondary, magnified back focal plane, which may be necessary in a practical setup, the displacement is increased by the magnification factor. In the next section, we will see that sub-wavelength focusing can lead to much larger deflections due to deviations from the paraxial field description and breakdown of the ponderomotive potential model.

However, we should remember that principally only the undiffracted part of the electron wave is affected by the laser focus. This part of the wave is, by definition, a plane wave without any information on the structure of the object.  
The deflection in the back focal plane, or translation in the object and image planes of the undiffracted wave does not lead to a loss of resolution in any way. Only the fact that, due to the finite size of the laser focus, also a part of the diffracted wave very close to the axis is affected, can lead to image degradation. As this part of the wave corresponds to small-angle scattering, the effect is an extra loss of phase contrast in the image at low spatial frequencies.

The change in axial velocity has no analog in light optics. It corresponds to a spatially localized frequency shift induced by the wave plate. However, in the quasiclassical theory it can be shown that this does not affect the phase difference between the diffracted and undiffracted paths of the particles at the detector position. For this, we start with Eq.~\ref{Eq:Phasedif_rel_f} and apply it to the trajectories from the end of the interaction region to the detector, where no laser field is present. Thus, the expression for the phase difference accumulated after the interaction region becomes (in the limit of small changes in axial velocity and small transverse velocities):
\begin{equation}
	\Delta\phi = \frac{-mc^2}{\hbar}\left[\int_{P_A} dt \frac{1}{\gamma}-\int_{P_N} dt \frac{1}{\gamma}\right]-\frac{\bm{p}_f}{\hbar}\cdot \Delta\bm{r} = \frac{-mc^2}{\hbar}\Delta\left[\frac{1}{\gamma}\right]t_\textrm{ID}-\frac{m\gamma v_z}{\hbar}\cdot \Delta v_z  t_\textrm{ID}
	\label{Eq:Phasedifn_rel_f}
\end{equation}
where $t_\textrm{ID}$ is the travel time from the interaction region to the detector. With $\Delta[1/\gamma] = \Delta[\sqrt{1-v_z^2/c^2}] = -\gamma v_z \Delta v_z/c^2 $, the first term in this expression cancels the second term, indicating that no phase shift accumulates after the interaction region even if the velocity (and hence the carrier frequency of the particle wave) is different in the two trajectories. 

This result is confirmed by (non-relativistic) wave packet calculations, in which a  wave packet that is slightly accelerated or decelerated by the interaction with a time-dependent ponderomotive potential representing the laser pulse interferes with a wave packet propagating at the original velocity. The net interference of the two packets (destructive or constructive) is only determined by the phase acquired during the interaction itself, and does not change any more when freely copropagating afterwards. 

This holds as long as the packets keep overlapping: as the group velocity of the two packets differs, eventually they do not overlap anymore. In practice, this does not constitute a problem. The energy spread of the electron beam sets a lower limit on the width in time of electron wave packets that we can use in any meaningful simulation of the electron microscope. With a spread of 1~eV on 200~keV beam energy, this limit is 0.7~fs. Any wave packet shorter in width needs a larger energy spread  to construct. Assuming a distance of 0.25~m from the back focal plane to the detector, shifted and unshifted wave packets with this minimal width start to separate at the detector if the group velocities are different by $>$100~m/s. Inspecting Table~\ref{Tab:Del_v_ana}, the predicted maximum change in axial velocity is below 3~m/s for a 100~fs laser pulse and below 1~m/s for a 300~fs pulse. 

Summarizing, the required phase shift can be achieved with pulsed laser parameters well within the range of (relatively) standard femtosecond laser systems. A tight focus is required for this. To minimize the laser power requirement and to maximize the spatial frequency range of the phase contrast image, the focus should be chosen as tight as possible. A longer wavelength reduces the power requirement as well.  However, this comes at the cost of increasing the minimum achievable waist size, which negatively affects the imaged spatial frequency range.
The required laser pulse energy scales linearly with the pulse duration. This favors short pulses. However, ideally the electron beam should be pulsed to a width of not more than a third of the laser pulse duration (and synchronized with an even smaller tolerance) in order to have an approximately constant phase shift of all electrons in the pulse. This is easier to achieve with longer laser pulses. Table~\ref{Tab:Energy_needed} covers a range of ``reasonable compromises'' for an effective phase plate.
Although the pulsed laser phase plate unavoidably causes a ``chirp'' in the axial velocity and a deflection sweep in the propagating direction of the laser, these do not seriously affect the operation of a phase contrast TEM, at least not for 30~fs electron pulses in combination with 100~fs laser pulses.

\section{Feasibility, simulations \label{Sec:Feasibiliy_sim}}
From the previous section it is obvious that we aim for a very small focus size. However, the smaller the waist size, the more important deviations from the paraxial approximation may become. To investigate this, we have performed a number of simulations using the approach outlined in Section~\ref{Sec:pondero_semi}.

\subsection{Phase shift \label{Subsec:sim_phase}}
\begin{figure}[h]
	\centering
	\includegraphics[width=150mm]{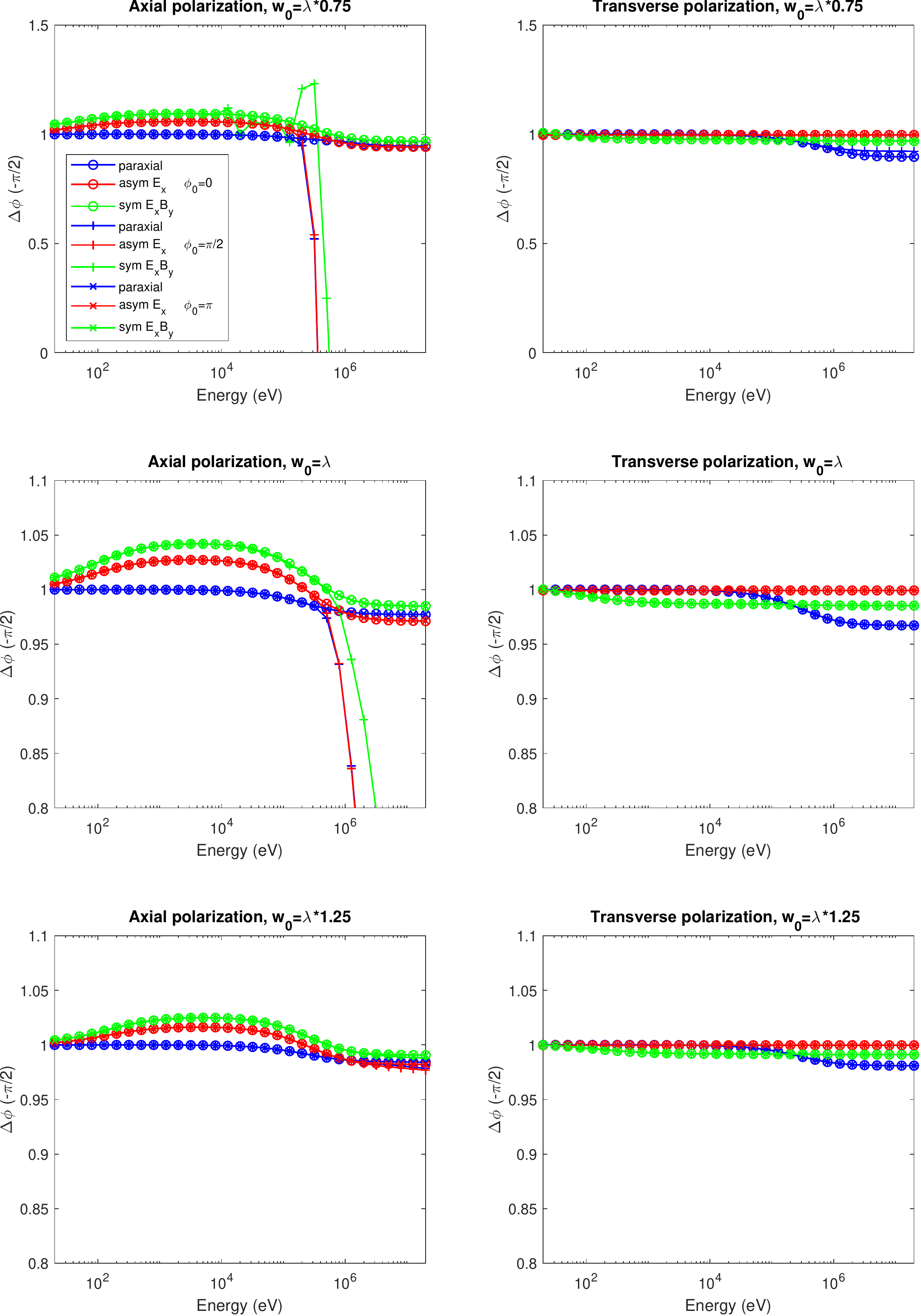}
	\caption{Phase shift in units of $-\pi/2$ as a function of theelectron energy from simulations. In the simulation, Gaussian laser pulses with wavelength $\lambda = 800\,\mathrm{nm}$, pulse duration $\tau = 100\,\mathrm{fs}$ and perfect synchronization ($t_0=0$) are assumed. The laser pulse energy is chosen for each simulation to give a phase shift of $-\pi/2$ according to the analytical model. In each frame, the results for three different field configurations and three different values for $\varphi_0$ are shown: blue symbols and lines represent the paraxial field model, red the angular spectrum model for a Gaussian distribution of $E_x$ (for axial polarization) or $E_y$ (for transverse polarization), and green the angular spectrum model for the symmetrized Gaussian $E_x$ -- Gaussian $B_y$ (resp.~Gaussian~$E_y$ -- Gaussian~$B_x$) distribution. Circles represent $\varphi_0=0$, $+$ symbols $\varphi_0=\pi/2$ and $\times$ symbols $\varphi_0=\pi$. For the phase shift, $\varphi_0=0$ and $\varphi_0=\pi$ results coincide.
	\label{Fig:Phase_shift_sim}}
\end{figure}
 
Fig.~\ref{Fig:Phase_shift_sim} shows the results for the calculated quasiclassical phase shift (in units of $-\pi/2$) based on calculated electron trajectories for a wide range of electron beam energy (20~eV -- 20~MeV), with a 100~fs laser pulse and 800~nm wavelength. The results are shown for three different waist sizes ($0.75 \lambda$ in the top frames, $\lambda$ in the middle frames and $1.25\lambda$ in the bottom frames) and for polarization along the electron beam axis (frames on the left) and perpendicular to the electron beam axis (frames on the right). In each frame, results for three different field models (the simple paraxial model, as well as the first and third of the models discussed in \ref{Subsec:beyond_parax}), and three values of the phase (0, $\pi/2$ and $\pi$) are shown. The pulse energy is adjusted for each simulation to give a phase shift of $-\pi/2$ in the analytical paraxial model (Eq.~\ref{Eq:Maxshift_rel}). Thus, the results of the analytical model correspond to a straight horizontal line at a value of 1 in each frame.

With the laser focused down to $0.75 \lambda$, the differences between the analytical model and the calculated quasiclassical phase shifts are large and depend strongly on the chosen field configuration and the polarization. At low energy the calculations with the simple paraxial field are independent of the polarization and in agreement with the analytical model, while the phase shifts calculated with the two physically correct (`Maxwell-compliant') field models are polarization dependent and, for axial polarization, considerably larger. This is largely explained by the distribution of the electric and magnetic energy densities in the focus, which for non-paraxial models are not rotationally symmetric and not equal. This deviation can still be incorporated in a ponderomotive potential model, using Eq.~\ref{Eq:pondero_pot} with the correct local electric field amplitude in the non-paraxial model. 

For higher energy, the calculated phase shift depends on the phase $\varphi_0$ of the field as well. This indicates a breakdown of the ponderomotive potential model at higher velocity, i.e., short interaction time. For axial polarization, where the electric field directly affects the kinetic energy of the electron, this effect is (for the symmetrized field model) already noticeable at 10~keV electron energy. At higher energy, fluctuations with the energy and eventually a reversal of the sign of the phase shift are visible. Above 200~keV, all models indicate that a usable phase plate cannot be achieved for axial polarization. For transverse polarization, no drastic effect is seen at any energy.

With less extreme focusing (focus size $\lambda$), no problems are seen in either polarization for energies below $\approx$ 800~keV. For $w_0=1.25\lambda$, the difference between the simulations and the analytical model is below $3\%$ for the whole energy range and only a very small dependence on $\phi_0$ is observed.

The values for the waist size in Fig.~\ref{Fig:Phase_shift_sim} are chosen to illustrate the relatively sharp onset of large deviations for higher electron energy around $w_0\approx \lambda$. For an even smaller waist size ($w_0=0.5\lambda$) these deviations become more extreme and start at lower electron energy, making a viable phase plate impossible for energies above 50~keV.

\clearpage

\subsection{Velocity changes \label{Subsec:sim_velo}}
\begin{figure}[h]
	\centering
	\includegraphics[width=77mm]{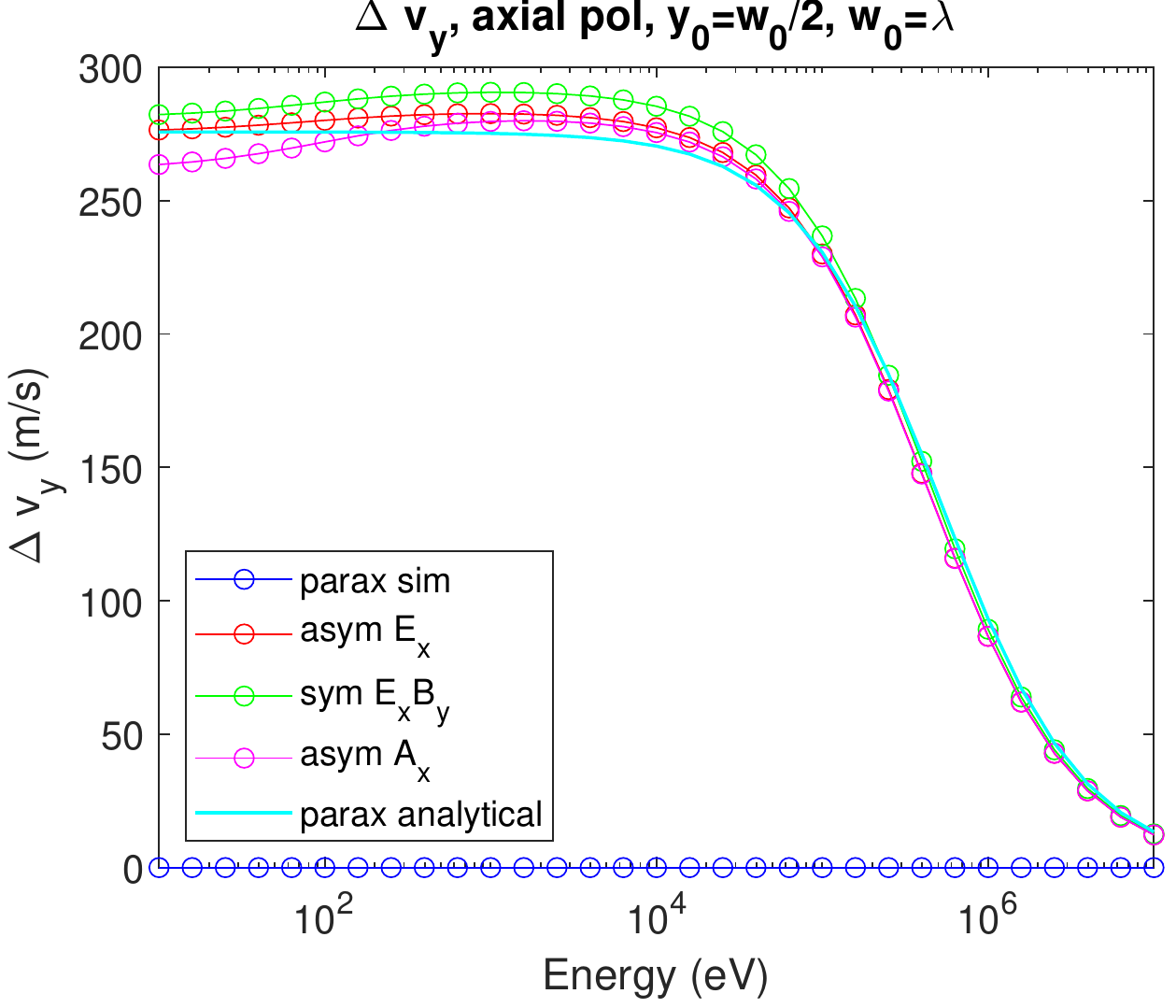}
	\caption{Change in axial velocity $v_y$ as a function of electron energy from simulations. Four different field configurations are shown: the paraxial approximation (blue circles and line), a Gaussian waist for $E_x$ (red), the symmetrized option (green, average of Gaussian $E_x$ and Gaussian $B_y$) and Gaussian $A_x$(magenta). The prediction of the analytical model with the effective ponderomotive potential of a paraxial Gaussian light field are represented by the cyan-colored curve. Waist size $w_0=\lambda$, light field phase $\varphi_0=0$. The other parameters are the same as in Fig.~\ref{Fig:Phase_shift_sim}.
	\label{Fig:Del_vy_h}}
\end{figure}

In the previous section, we have shown that the quasiclassical phase shift as derived from classical trajectories in a paraxial field generally agrees well with the values derived from the trajectories in more correct field approximations, as well as with the analytic result based on the ponderomotive potential in a paraxial field. However, we do not expect the same to be true for the calculated velocity changes of the electron. To illustrate this, Fig.~\ref{Fig:Del_vy_h} shows the calculated change in velocity along the $y$--direction for an electron that does not pass through the center of the laser beam (with axial polarization), but passes at an offset $y_0=w_0/2$ in the $y$--direction. The ponderomotive potential model (the cyan-colored line in the figure) results in a velocity change in the $y$--direction, which is essentially a result of the repulsion by the central maximum of the potential.

For this waist size, all trajectory calculations using physically correct field models agree well with the ponderomotive potential model (for which the potential is derived using the paraxial approximation of the field). However, using the same paraxial approximation for the field in trajectory calculations leaves the velocity in the $y$--direction unchanged. 

This is immediately clear even without actually calculating the trajectories: with the electric field along the $x$--axis, the magnetic field along the $y$--axis and the electron initially traveling in the $x$--direction, the total Lorentz force on the electron always remains in the $xz$--plane. 

The calculated velocity changes using the $E_x$ and symmetrized models, under the same conditions as the phase shift results in Fig.~\ref{Fig:Phase_shift_sim}, are shown and discussed in Appendix~\ref{Appendix:Velocity_sim}. The overall conclusion that can be drawn from these results coincides with the one from the previous section: No problem should arise unless focusing down to a sub-wavelength waist size. If pushing the limit in focus size is pursued, choosing transverse polarization of the light field is preferable in order to minimize velocity changes.

\section{Conclusions}
The focused beam of a femtosecond laser can be realistically used as an effective, simple, and variable pulsed phase plate for a TEM. The required pulse energy for the optimal phase retardation of $\pi/2$ for a Zernike-type phase microscope is well within the range offered by off-the-shelve, high repetition rate laser systems. Extending the phase retardation to a maximum of $2\pi$, allowing a complete reconstruction of the transmitted electron wavefront, seems feasible as well. 

The simple picture of an effective ponderomotive potential with a paraxial Gaussian laser pulse predicts that the laser focus should be as tight as possible to minimize power requirements and to improve the image contrast at low spatial frequencies. However, the tighter the focus is, the less valid the paraxial approximation. 
A more accurate description of the light field is necessary. We have developed a model for a pulsed laser phase plate in which relativistic electron trajectories are calculated in the field of a light beam with an arbitrarily small focus in a choice of exact, non-paraxial descriptions.  A quasiclassical path integral approach is used to derive the electron phase shift from these trajectories. Changes in the final velocity of the electron can be translated into imaging errors.

The results indicate that, for 800~nm wavelength, 100~fs laser pulse duration, and 200~keV electrons, a focus size down to around one wavelength is acceptable. Below this value, deviations from the paraxial ponderomotive potential model may cause image degradation. At higher electron energy, polarization of the light along the electron beam axis leads to a strong dependence of the calculated final phase shift on the phase of the light field at the time the electron passes the center of the interaction region. As this time is different for each electron, this leads to a large spread in the phase shift. This problem does not occur for a light field polarized perpendicular to the electron beam axis, or with a slightly larger focus size of 1.25~$\lambda$.

In combination with RF electron bunch chopping and compression techniques, the phase plate can be used to construct a pulsed, high repetition rate phase contrast microscope with an acceptable duty cycle. First chopping the electron beam in bunches and subsequently compressing these bunches in the time domain enables short electron pulses with an overall duty cycle of up to 30\%~\cite{Toonen2019}. However, time domain compression unavoidably introduces an energy spread in the electron beam. 

As an example, chopping a 200~keV beam with a 3~GHz RF cavity to 100~ps width pulses (i.e., a duty cycle of 30\%) and compressing these pulses to the sub-picosecond regime with a single compression cavity with 1~m focal length introduces an electron energy spread of $\approx 7\%$, fully unrealistic for an electron microscope because of the resulting chromatic aberration. However, with a more elaborate multi-stage compression scheme, this spread can be significantly reduced. The ultimate limit is set by phase space area conservation, with the energy spread increasing inversely proportional to the pulse duration. For a microscope with 0.5~eV energy spread, compressing a 100~ps pulse to 300~fs will then lead to 167~eV or 0.084\% final energy spread.

Without compression, chopping to 100~fs directly would lead to a duty cycle of $3\times 10^{-4}$ at 3~GHz without increasing the energy spread. Between these extremes, a compromise can be sought between duty cycle and energy spread.   

The calculational approach developed for the ponderomotive phase plate will be used to further optimize the design of the phase plate. Furthermore, it also forms a flexible and powerful modeling tool for the electron-light interaction with a tightly focused beam in a variety of configurations.

\newpage 

\appendix
\section{Velocity changes from simulations\label{Appendix:Velocity_sim}}
\begin{figure}[h]
	\centering
	\includegraphics[width=150mm]{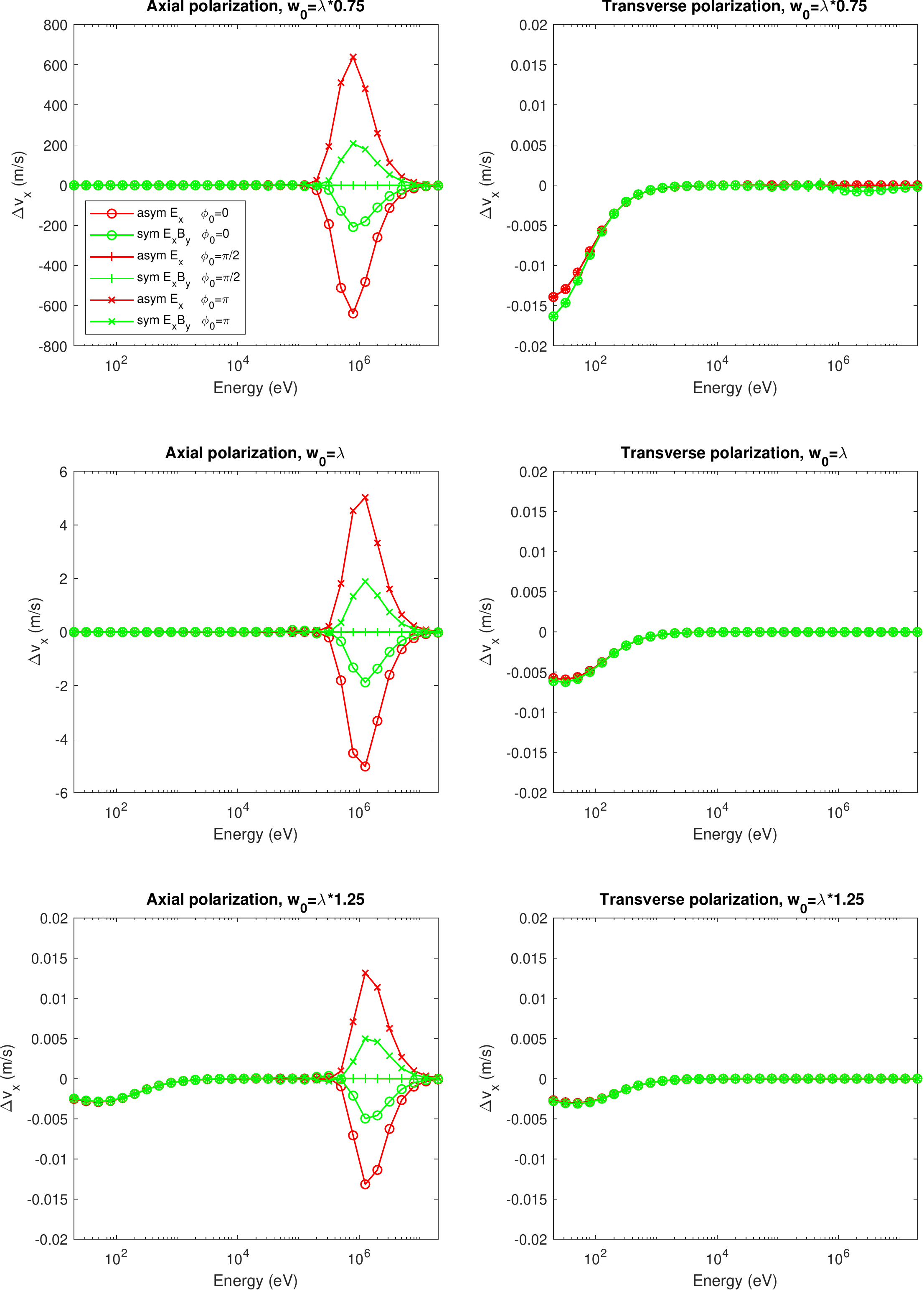}
	\caption{Change in axial velocity $v_x$ as a function of electron energy from simulations. Laser parameters and identification of curves the same as in Fig.~\ref{Fig:Phase_shift_sim}.
		\label{Fig:Del_vx}}
\end{figure}

\begin{figure}[h]
	\centering
	\includegraphics[width=150mm]{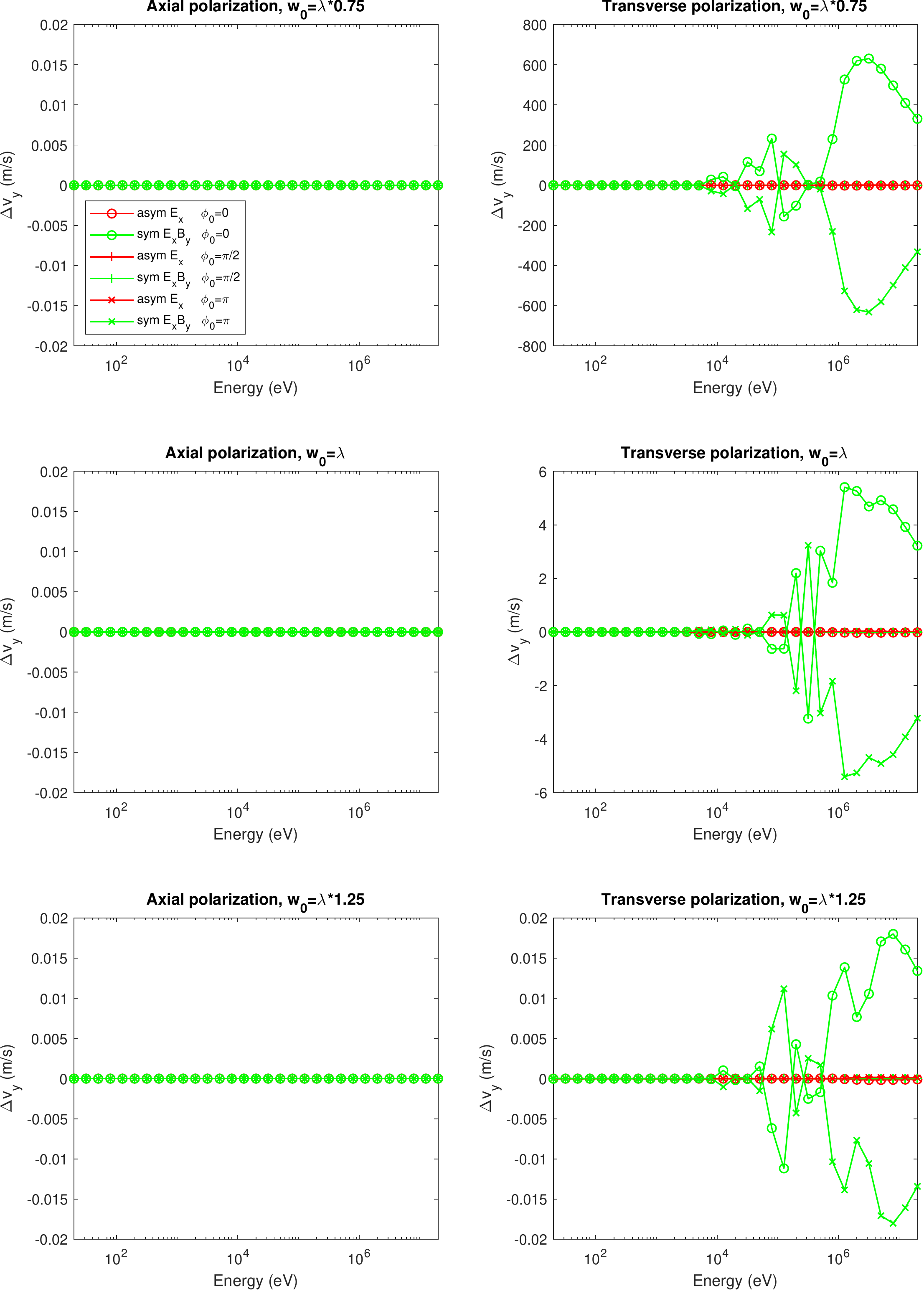}
	\caption{Change in transverse velocity $v_y$ as a function of electron energy from simulations.  Laser parameters and identification of curves the same as in Fig.~\ref{Fig:Phase_shift_sim}.
    \label{Fig:Del_vy}}
\end{figure}

\begin{figure}[h]
	\centering
	\includegraphics[width=150mm]{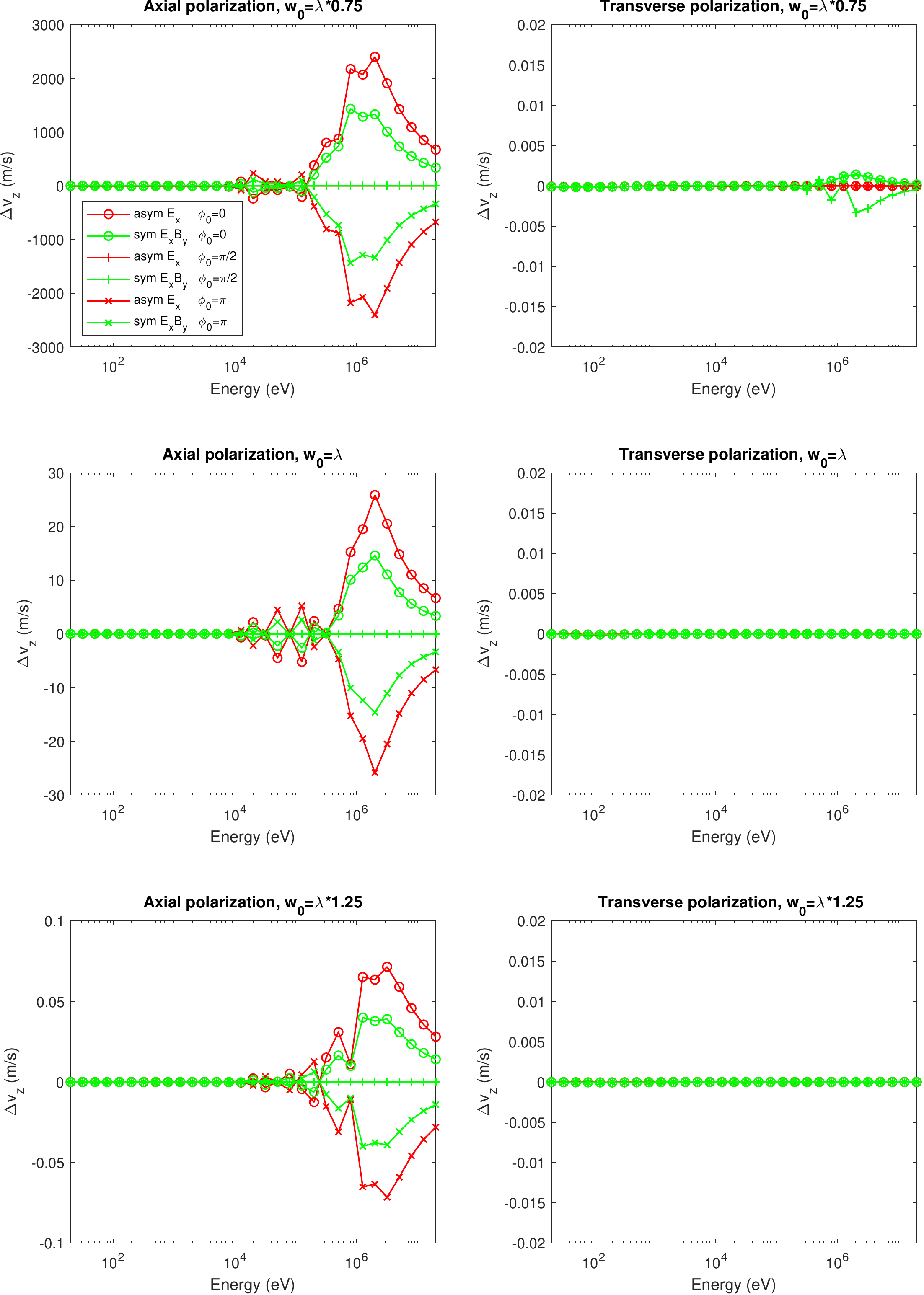}
	\caption{Change in transverse velocity $v_z$ as a function of electron energy from simulations.  Laser parameters and identification of curves the same as in Fig.~\ref{Fig:Phase_shift_sim}.
	\label{Fig:Del_vz}}
\end{figure}

In Figures~\ref{Fig:Del_vx}--\ref{Fig:Del_vz}, the changes in $v_x$, $v_y$ and $v_z$ are shown for the same parameters as in Fig.~\ref{Fig:Phase_shift_sim}. As $t_0=0$ and $y_0=0$, the analytical model predicts no velocity changes in any direction. The depicted velocity changes are purely the result of the detailed microscopic motion of the electron in the laser field.

Inspecting these figures, we again see a very strong dependence on the focusing parameter of the laser beam. For a waist $w_0=1.25\lambda$, no significant changes in the velocity are seen at any electron energy. For $w_0=\lambda$, velocity changes in the $y$-- and $z$--direction (depending on the polarization, phase, and the field model chosen) are observed for energies above 10~keV. With transverse polarization, the absolute value of the velocity changes only exceeds 2~m/s in any direction for a beam energy $>$200~keV. At $w_0=0.75\lambda$, velocity changes of up to $600$~m/s in the axial direction ($x$), up to $2500$~m/s in the direction of the laser beam propagation ($z$), and up to $600$~m/s in the direction perpendicular to both ($y$). Although the largest changes are seen for energies in the MeV range, at 200~keV they are still sizable.

The trajectory calculations demonstrate the influence on the final electron velocity components of the deviations from the analytical paraxial model. Based on these results, using a sub-wavelength waist size is not advisable. Using $w_0=\lambda$ (or slightly larger, just to be on the safe side) does not lead to significant imaging errors and is therefore acceptable.
\clearpage

\bibliography{Ponderomotive_Phase_Plate2.bib}

\end{document}